\documentclass[useAMS,usenatbib]{mn2e}
\usepackage{graphicx,color,epsfig}

\newcommand{\be}{\begin{equation}}
\newcommand{\ee}{\end{equation}}

\newcommand{\apj}{ApJ}

\newcommand{\mnras}{MNRAS}
\newcommand{\aap}{A\&A}
\newcommand{\araa}{ARA\&A}
\newcommand{\apjl}{ApJL}
\newcommand{\aj}{AJ}
\newcommand{\nat}{Nature}

\def\ltsima{$\; \buildrel < \over \sim \;$}
\def\simlt{\lower.5ex\hbox{\ltsima}}
\def\gtsima{$\; \buildrel > \over \sim \;$}
\def\simgt{\lower.5ex\hbox{\gtsima}}
\def\sgra{Sgr~A$^*$}

\newcommand\ledd{{L}_{\rm Edd}}

\def\msun{{\,{\rm M}_\odot}}

\newcommand\mbh{{\,{\rm M}_{\rm bh}}}

\newcommand\mearth{{\,{\rm M}_{\oplus}}}

\def\del#1{{}}

 \title[SMBHs and Super-Oort clouds]{Are SMBHs shrouded by
   ``super-Oort'' clouds of comets and asteroids?}

\author[S.~Nayakshin, S. Sazonov and R. Sunyaev]
{\parbox{18cm}{Sergei Nayakshin$^{1,3,*}$, Sergey
    Sazonov$^{2,3}$ and Rashid Sunyaev$^{2,3}$}\vspace{0.3cm}\\
$^1$Dept. of Physics \& Astronomy, University of Leicester, Leicester, LE1
  7RH, UK\\
$^2$IKI, Moscow, Russia\\
$^3$Max-Planck-Institut f\"{u}r Astrophysik, Karl-Schwarzschild-Stra\ss{}e 1,
85741 Garching bei M\"{u}nchen, Germany}

\begin{document}

\maketitle

\begin{abstract}
Recent decade saw a dramatic confirmation that an in situ star
formation is possible inside the inner parsec of the Milky Way. Here
we suggest that giant planets, solid terrestrial-like planets, comets
and asteroids may also form in these environments, and that this may
have observational implications for Active Galactic Nuclei (AGN).
Like in debris discs around main sequence stars, collisions of large
solid objects should initiate strong fragmentation cascades. The
smallest particles in such a cascade -- the microscopic dust -- may
provide a significant opacity. We put a number of observational and
physical constraints on AGN obscuring torii resulting from such
fragmentation cascades. We find that torii fed by fragmenting
asteroids disappear at both low and high AGN luminosities.  At high
luminosities, $L\sim \ledd$, where $\ledd$ is the Eddington limit, the
AGN radiation pressure blows out the microscopic dust too rapidly. At
low luminosities, on the other hand, the AGN discs may avoid
gravitational fragmentation into stars and solids. We also note that
these fragmentation cascades may be responsible for astrophysically
``large'' dust particles of $\simgt \mu$m sizes that were postulated
by some authors to explain unusual absorption properties of the AGN
torii.
\end{abstract}

\renewcommand{\thefootnote}{\fnsymbol{footnote}}
\footnotetext[1]{E-mail: {\tt Sergei.Nayakshin at astro.le.ac.uk}}

\section{Introduction}
\label{intro}

Active Galactic Nuclei (AGN) are galactic centres powered by Super-Massive
Black Holes (SMBH) growing by accretion of gas (REF). The specific angular
momentum of gas in a galaxy is too large to be accreted onto the SMBH
directly. An accretion disc \citep{Shakura73} is required to transfer the
angular momentum outward and mass in.  The disc is massive and cool at large
distances from the SMBH. For this very reason, AGN accretion discs are found
to be gravitationally unstable and should collapse into stars beyond $\sim
0.01-0.1$ pc\citep{Paczynski78,Kolykhalov80,Lin87,Collin99,Goodman03}. The
fragmentation process has been confirmed in numerical simulations
\citep{NayakshinEtal07,AlexanderEtal08}, and appears to be the only reasonable
explanation for the two discs of young stars in the central $\sim 0.5$ pc of
our Galaxy \citep{Levin03,GenzelEtal03,NC05,PaumardEtal06}.

Stars forming elsewhere in the Galaxy frequently, and perhaps always,
form with planets and debris discs. In this paper we shall argue that
stars forming in AGN discs should also come with their own planets,
both giant and terrestrial-like, and should also have solid ``debris''
around them --asteroids and comets. We show that planets and asteroids
formed in the outer fringes of the proto-stellar disc around the
parent star are stripped away by perturbations from close passages of
stars in the AGN disc. 

Released from their host stars, these solids and planets orbit the SMBH
independently. Since the velocity kick required to unbind them from the host
is in km/s range, whereas the star's orbital velocity around the SMBH is
$\simgt 1000$ km/s, orbits of the solids are initially only slightly different
from that of their hosts.  AGN gas discs are expected to be very geometrically
thin \citep[e.g.,][]{NC05}, and if they always lay in the same plane (e.g.,
the disc galaxy's mid-plane) then the resulting distribution of solids would
be quite thin and planar as well.

However, there is no particularly compelling reason for a single-plane mode of
accretion in AGN as the inner parsec is such a tiny region compared with the
rest of the bulge \citep{NK07}, and chaotically-oriented accretion may be much
more likely \citep[e.g.,][]{KingPringle06}. One physically plausible way to
obtain randomly oriented inflows in AGN is turbulence, and more generally,
random motions, excited by the star formation feedback in the bulge of the
host \citep[cf. simulations of][]{HobbsEtal11}.

Furthermore, there is one well documented observational example of such a
geometrically thick distribution of {\em stars} in the centre of our
Galaxy. The two young stellar discs in the Galactic Centre are inclined to
each other at a large angle, and the discs are possibly strongly
warped \citep{BartkoEtal09}. While the nature and even the existence of the
second disc is debated \citep{LuEtal09}, it is plainly clear that the system
of young stars in the centre of our Galaxy is very thick kinematically. A
natural way to create such a complex distribution of stellar orbits is to have
a violent collision of a Giant Molecular Cloud (GMC) with another cloud or a
pre-existing gas disc \citep{HobbsNayakshin09}, or have multiple GMC
deposition events \citep{BonnellRice08,AligEtal11}.

In an AGN that is fed by repetitive randomly oriented ``feeding episodes'', we
expect many stellar rings forming over time at different angles to the galaxy
plane. The distribution of solids torn away from their parent stars should
thus be similarly multi-plane. Due to gravitational precession of different
discs with respect to one another \citep{NayakshinEtal06}, the system should
evolve into a roughly quasi-spherical configuration or a thick torus if there
is a preferred direction like the galaxy plane.  We refer to this torus of
solid bodies as a super-Oort cloud of SMBH due to an obvious analogy with the
Solar System's Oort cloud (interestingly, the physical sizes of the two clouds
would be roughly the same).

The total mass in the solids, compared to that in the gas or  the stars
within the inner parsec, should be tiny simply because H and He far outweigh
metals in a gas of Solar composition. Accordingly, we do not expect any
significant effects of the solids on the dynamics of stars or gas in the
central parsec. However, the smallest solids -- microscopic dust particles --
have very large absorption cross section per unit mass and may be important
for the observational appearance of AGN.

We now note an observational analogy. Dust around main sequence stars absorbs
the incident radiation and re-radiates it into the IR and NIR
wavelengths. These dusty reservoirs around ``normal stars'' are called debris
discs because they are believed to be dominated by large ($\sim 100$ km or so)
solid bodies which are the remnants of the planet formation process. Debris
discs around stars are optically thin \citep{Wyatt08}. Due to this,
microscopic dust is continuously driven away by the radiation pressure from
the star and thus needs to be replenished. The replenishment occurs via a
top-down cascade of larger solid bodies colliding and fragmenting on ever
smaller objects. This general picture is supported by the observations of such
collisional cascades in our own Solar System, in the Kuiper and the asteroid
belts. Some of the debris discs around nearby stars have been directly imaged
\citep[e.g.,][]{SmithEtal09}.

The relative velocities of solids in the Super-Oort cloud around SMBHs
envisaged here are as large as hundreds to thousands km s$^{-1}$, compared
with $\simlt$ km s$^{-1}$ for debris discs around stars in the Solar
neighborhood. This should fuel a powerful fragmentation cascade that may in
principle lead to an optically thick veil of dust, perhaps contributing to the
observed but still poorly understood AGN obscuration (cf. \S
\ref{sec:unification} for literature on this). On the other hand, dust
particles could be quickly blown away, could collapse into a geometrically
thin disc, or be sublimated by the AGN radiation; the cascade may also become
depleted ``too soon''. 

The purpose of our paper is to put physical constraints on the fragmentation
cascade of solid bodies, such as asteroids and comets, around a SMBH to
determine its observational significance for the AGN phenomenon.  We start in
\S 2 with arguments for feasibility of planet formation near AGN. In \S
\ref{sec:fragmentation} we estimate time scales for catastrophic collisions of
large bodies feeding the cascade to smaller scales, and in \S \ref{sec:super}
observations of AGN obscuration are used to constrain the population of small
$\sim \mu$m size grains. To simplify the estimates, we assume that large
solids bodies (comets, asteroids and Moon-sized objects) dominate the mass in
the fragmentation cascade, whereas the small grains provide all the opacity
(obscuration). In \S \ref{sec:discussion} a general discussion of the
constraints on and the implications of our model is given.  We note that our
work does not exclude the ``conventional'' gas-rich torus models for AGN torii
(see the references in \S \ref{sec:unification}). It is possible that AGN
obscuration is achieved by a variety of means in different sources or at
different epochs of AGN evolution.

\section{Planets, comets and asteroids near a SMBH}\label{sec:planets} 

\subsection{Birth of planets}\label{sec:birth}

Can planets and/or asteroids form in the protoplanetary discs around
their parent stars that orbit a SMBH at velocities as large as $\sim
1000$ km~s$^{-1}$? The theory of planet formation is in itself an
active area of research with widely differing opinions on how the
process works, thus the answer on the question is model
dependent. Nevertheless, we shall argue that the answer on the
question above is probably ``yes''.

One physical constraint on the planet formation process around a star in the
AGN environment is that the required proto-planetary
disc must fit within the Hills radius, $R_H$, of the parent star, or else the
disc would be truncated by the tidal forces from the SMBH. For the parent star
with mass $M_* = 1\msun$, orbiting the black hole of mass $\mbh = 10^8\msun \,
M_8$, a distance $R$ away, the Hill's radius is
\begin{equation}
R_H = R \left( {M_*\over 3 \mbh} \right)^{1/3} = 300\; \hbox{AU}\; R_{\rm pc}
M_8^{-1/3}\;,
\label{rh}
\end{equation}
where $R_{pc} = R/(1$pc).  For solids formed closer than $\sim 100$ AU to
their host stars, this constraint is satisfied unless $R_{\rm pc} \ll 1$.

\subsubsection{By Core Accretion}

The earliest versions of the Core Accretion (CA) model for planet formation
were based on the observations of the Solar system only
\citep[e.g.,][]{Safronov69}, and suggested a rather slow formation of
terrestrial planets from ``planetesimals'', e.g., solid bodies with size of a
few km \citep[for a review see][]{Wetherill90}. Solids of these comparatively
small sizes have surface escape velocities of the order of $\sim$
m~s$^{-1}$. Collisions at velocities exceeding that would be shattering,
therefore the initial growth is assumed to occur in razor-thin discs with very
small velocity dispersions. This mode of rocky planet formation would not be
able to build any large solid bodies near an AGN due to the rapid
collisional destruction of such bodies in collisions at $\sim 1000$
km~s$^{-1}$ (cf. equation \ref{tcoll3} below).  In other words, even if the
planetesimals were born in the protoplanetary discs near the stars, no large
solid objects would be able to grow from them and the planetesimals would be
very quickly destroyed themselves in fragmenting collisions.

However, more recent work indicates that a more rapid planetesimal
growth may happen due to turbulence and instabilities in the gas-dust
disc \citep[e.g.,][]{YoudinGoodman05,JohansenEtal07}. In these models,
small rocks are first concentrated into regions shaped by the motions
of the gas flow around them, and then undergo a gravitational collapse
to form larger solids.  Numerical simulations suggest that growth of
solids up to minor planet sizes may occur as rapidly as within $\sim
10^3-10^4$ yrs in this setting \citep{JohansenEtal07}. It appears that
this mode of planetesimal and rocky core formation may be resilient
enough to the tidal and radiation field challenges of central parsecs
of galactic nuclei.

\subsubsection{By gravitational disc fragmentation}

Observational evidence is accumulating for planets that probably did
not form according to the CA scenario.  The giant planets observed at
many tens and hundreds of AU from their parent stars \citep[see
  references and discussion in][]{Boley09,MCR10} could not have their
cores assembled on reasonably short time scales
\citep{Rafikov11}. Therefore, these planets are believed to have
formed by the gravitational disc instability (GI) in the outer fringes
of proto-stellar accretion discs
\citep{MayerEtal04,Rice05,DurisenEtal07,SW08,Boley09}. This is a
miniature version of the star formation process near an AGN, with
planets born around individual stars.

An important modification of the GI theory allows formation of
terrestrial-like planets and all sort of smaller solids such as the
asteroids, breaking the quarter-century held belief that rocky planets
can be only formed by the CA.  \cite{BoleyEtal10,Nayakshin10c} have
recently showed that radial migration of the proto giant planets may
pave the way to formation of terrestrial planets. In these models,
which we call here ``Tidal Downsizing'' (TD), a ``large'' $R_{\rm
  disc} \sim 100$ AU gas-dust disc fragments onto many
self-gravitating clumps. These clumps migrate inward rapidly. Dust
grains within the clumps grow and sediment to the centre of the clumps
into massive solid cores -- the proto terrestrial planets. The gaseous
envelopes of the clumps may be removed by the tides from the parent
star or by stellar irradiation and later accreted onto the star. Only
the solid core (the future rocky proto-planet) is left behind as its
material density is far higher than that of the gaseous
embryo. Rotation, not taken into account in this simplest picture,
prevents segregation of all of the solid material into the centre, and
may lead to formation of planetary satellites such as the Moon
\citep{Nayakshin10d}, and other smaller solid debris. Disruption of
the gaseous envelope then releases the solid debris into the
surroundings \citep[cf. Figs. 9 and 10 in][for small grains disrupted
  together with the gas; Cha and Nayakshin, in preparation, show that
  large asteroids are also unbound from the proto-planet except for
  those inside its Hill's radius]{ChaNayakshin10}.

Formation of solid bodies in the TD framework is a very rapid process, taking
$\sim 10^3-10^4$ yrs \citep{Nayakshin10a,Nayakshin10b,ChaNayakshin10}. Thus
solids would be in place as rapidly as the star formation starts in the AGN
disc in this model. 

\subsubsection{Starless planets in AGN discs}

\cite{Shlosman89} noted that not only stars but giant gas planets can be born
directly in AGN discs (not around individual stars). Similarly,
\cite{Nayakshin06a} found that the Toomre mass of the marginally
self-gravitationally unstable disc may permit formation of objects in the
planetary mass regime (cf. his Fig. 1), although the final outcome of disc
gravitational fragmentation depends on whether the massive planets/stars
accrete more mass or not.

It is now understood that planets formed by the GI can hatch solid cores
\citep{Boss98,BoleyEtal10,Nayakshin10b}. If these are then stripped of their
gas envelope by tidal forces, irradiation, passages through the disc \citep[as
  argued for stars by][]{GoodmanTan04} or close interactions, then these cores
are also liable to participate in the collisional fragmentation cascade
discussed below, provided that the impactors are large enough.

\subsection{Decoupling from the parent star}\label{sec:decoupling}

As in the case of the Oort cloud in the Solar System, one can expect
that smaller solids such as asteroids formed in the proto-planetary
disc are scattered by planets onto larger orbits, and some are lost
from the Hill's sphere of the parent star altogether
\citep[cf. references in ][]{Fernandez97}. Besides the Oort cloud
itself, the most convincing evidence for the importance of this
process comes from observations of ``freely floating'' giant planets
\citep{SumiEtal11} that may number as many as two per a main sequence
star in the Milky Way. These planets were probably expelled from their
parent systems by close encounters with even more massive
bodies. Anything less massive than a giant planet would be even more
likely to suffer a similar fate, then.

Close passages of stars can also strip away solid bodies in the outer
reaches of the system. In the impulse approximation, the momentum
passed to a solid body by a star passing with a relative velocity
$v_{\rm rel}$ scales as $\propto v_{\rm rel}^{-1}$ \citep[cf. \S 1.2.1
  of][]{BT08}. Therefore, only encounters with small relative
velocities are important in stripping the solids from the parent
stars. During the birth of the stars and planets in the AGN accretion
disc, the relative velocity is expected to be $\sim (H/R) v_K \simlt $
tens of km s$^{-1}$ \citep{NC05}. Consulting Fig. 2 of \cite{ZT04}, we
see that solids in the outer tens of AU disc should be stripped by
just a few stellar passages within a few hundred AU of the parent
star. Such passages are frequent inside the AGN star-forming
discs. Indeed, the gas disc vertical scale height is expected to be
$H\sim 0.01 R = 200 $ AU at $R = 0.1$ pc \citep{NC05}. The mean
stellar separations may be expected to be of the order $H$
\citep[cf. related estimates for gas clump-clump collisions
  in][]{Levin07}.

We therefore conclude that solids born in the gas-dust discs in the outermost
$\sim$ tens of AU from their parent stars are vulnerable to external
perturbations releasing them into independent orbits around the SMBH. Solids
born in the inner ten or so AU are much harder to perturb out of the grips of
their host stars. These are more likely to be excited onto eccentric orbits
within the Hill's radius of the star. However, these star-bound solids do
participate in the global fragmentation cascade as they are hit by
``external'' solids at the same rate as solids on their own independent
orbits. Therefore, within our approximate exploratory model, we can consider
both populations in the same manner.

\section{Fragmentation cascade}\label{sec:fragmentation}

The problem we wish to study is not amenable to an exact analytic study. Our
goal here is to place order of magnitude constraints on the picture being
proposed keeping the arguments as transparent as possible. In the context of
fragmenting asteroids and debris discs around stars, one frequently assumes a
quasi steady state fragmentation cascade of solids to form
\citep{Dohnanyi69,Wyatt08}. In this case solids of a given size $a$ are
removed by fragmentation at the same rate as solids of the same size arrive
due to fragmentation of larger bodies.

For the problem at hand, such a steady state might exist at
intermediate sizes only because the smallest grains are strongly
affected by aerodynamic gas drag, whereas the largest objects may not
have had enough time to experience a catastrophic (shattering)
collision.  Therefore, we chose to consider two populations of solids
separately. The largest solids have diameter $D$ and dominate the
system in terms of mass. The rate of their fragmentation determines
the rate of dust production. The small grains considered in \S
\ref{sec:small}, e.g., the dust, dominate the absorption opacity of
the system.

\subsection{Fragmentation of large bodies}\label{sec:large}

Consider large solids in Keplerian orbits around the SMBH with
semi-major axis $\sim R$ and number density $n_D$. One can estimate
the self-collision time scale for these large solids immediately, and
that is likely to be quite long. However, every one such collision may
create thousands and millions of smaller objects. Many of these
fragments may be large enough to split the large bodies (see equation
\ref{d1} below). Furthermore, it is not realistic to assume that no
small bodies are born during the star and planet formation in AGN
discs (cf. the arguments in the end of this section).

Therefore, we shall estimate the time scale on which the large bodies
$D$ are destroyed differently. We assume that one way or another, a
fragmentation cascade develops. We take a power-law form for the
cascade with $\Delta n_d = n_D^0 (d/D)^{-q} \Delta d$ giving the
number density of asteroids with diameter between $d$ and $d+\Delta
d$. Steady-state fragmentation cascades results in $3.5 \leq q \leq 4$
\citep{Dohnanyi69,KennedyWyatt10}. We estimate the number of asteroids
of size $d$ as
\begin{equation}
n_d \sim d\left({\Delta n_d\over \Delta d}\right) \sim n_D (d/D)^{-i}\;,
\label{ndsmall}
\end{equation}
where $i = q-1$ and $n_D = n_D^0/D$.

The rate at which solids are ground is determined by catastrophic collisions,
e.g., collisions resulting in fragmentation of the larger body. The size $d$
of the impactor in a catastrophic collision is determined from the condition
that its kinetic energy relative to the asteroid $D$ is equal to the binding
energy of the latter, $\sim G [(\pi/6)\rho_a D^3]^2/D$. The relative velocity
of the solids before the collision is
\begin{equation}
v_{\rm rel} = \delta v_K \approx
1000 \; \hbox{km s}^{-1} \; \delta M_8 R_{pc}^{-1/2}\;,
\label{vrel}
\end{equation}
where $\delta \simlt 1$ is a dimensionless parameter. This velocity is clearly
far higher than the range of collision velocities studied in the context of
asteroids around stars in ``normal'' Galactic environment.

The minimum size of the impactor that would split the asteroid of diameter $D$
is
\begin{equation}
d = \left( {\pi G \rho_a \over 3 v_{\rm rel}^2}\right)^{1/3} D^{5/3}\;.
\label{d1}
\end{equation}
Numerically,
\begin{equation}
{d \over D} \approx 4.6 \times 10^{-3} D_8^{2/3} v_{8}^{-2/3}\;,
\label{d2}
\end{equation}
where $D_8 = D/1000$km and $v_8$ is $v_{\rm rel}$ in units of 1000 km
s$^{-1}$.  This shows that a 1000 km size asteroid can be split by a
projectile of only $\sim 5$~km across if the impact occurs at $v_{\rm
  rel} \sim 1000$ km s$^{-1}$.

The catastrophic collision time scale can be estimated as $t_{\rm
  coll} = [n_d v_{\rm rel} \pi (D/2)^2]^{-1}$ which results in
\begin{equation}
t_{\rm coll} \approx \left( G \rho_a \right)^{i/3}  n_D^{-1} v_{\rm rel}^{-2i/3-1} D^{5i/3-2}
\label{tcoll1}
\end{equation}

In order to constrain this further, let us relate $n_D$ to the total mass of
solids, $M_Z$, occupying volume $\sim (2\pi/3) R^3$ (i.e., half of spherical
volume of radius $R$ as seen from the SMBH). As the mass of one asteroid is
$(\pi/6)\rho_a D^3$, we have $n_D = (3/\pi)^2 M_Z/(\rho_a D^3 R^3)\sim
M_Z/(\rho_a D^3 R^3) $. Hence we estimate 
\begin{equation}
t_{\rm coll} \approx \rho_a^{i/3+1} G^{-1/2}R^{i/3+7/2}
D^{2i/3+1}M_Z^{-1}\mbh^{-i/3-1/2}\;.
\label{tcoll2}
\end{equation}
For $\rho_a = 2$ g cm$^{-3}$, and $i = 2.5$, 
\begin{equation}
t_{\rm coll} \approx 5 \times 10^6 \,\hbox{yr}\;
R_{pc}^{13/3}D_8^{8/3}M_{Z3}^{-1} M_8^{-4/3}\;,
\label{tcoll3}
\end{equation}
Here $M_{Z3} = M_Z/10^3\msun$.  For reference, setting $i=3$ results in
$t_{\rm coll} \approx 0.4$ Myr for the same nominal parameters as above.

Equation (\ref{tcoll3}) demonstrates that very large solid bodies,
e.g., of a planetary satellite size, such as the Moon, are able to feed
the cascade for interestingly long time scales. Bodies of smaller
sizes would be ground down too quickly to be observationally
important.

``Primordial'' solid formation (e.g., concurrent with the star
formation itself) in normal Galactic conditions creates a distribution
of solid bodies which is currently not well known
\citep[see][]{SW11}. Consider now the modifications to our picture
resulting from the large bodies themselves having a range of sizes
rather than a fixed size. Because the collision time is a very strong
function of the asteroid size $D$, equation \ref{tcoll3} shows that
smaller bodies will be shattered very quickly. On the contrary, the
largest bodies, e.g., the size of the Earth, are fragmented on very
long time scales ($10^9$ yrs or more). At any given time $t$, counted
from the last star and planet formation event, the fragmentation
cascade is fed by asteroids of size $D(t)$ such that $t_{\rm coll}(D)
= t$. Most of the asteroids larger than $D(t)$ have not yet had time
to fragment by the time $t$. $M_Z$ in this case should be understood
as the mass of the asteroids with diameter $D=D(t)$.

\section{AGN absorbers and small grains in the super-Oort clouds}\label{sec:super}

As stated in the Introduction, in this paper we explore whether AGN absorbers
may be more massive cousins of stellar debris discs, made up of solids from
$\sim$ millions of debris discs around individual stars. Having discussed
the fragmentation cascade of the largest solids, we now assume that the end
result of that is a population of dust particles able to absorb and re-radiate
the AGN radiation. In the spirit of an exploratory model, the population of
the smallest grains is described by grains of a single radius, $a$, which is
taken to be of $\mu m$ size. We refer to these grains as ``small'' in
comparison to the larger solid bodies we consider, even though they may be
somewhat large by the standards of Galactic grains, where $a\simlt 0.25 \mu$m
\citep{MathisEtal77}. As for debris discs \citep{Wyatt08}, we assume that the
small grains completely dominate the opacity of the absorber, whereas the mass
is dominated by the large bodies.

\subsection{Observations and models of AGN absorbers}\label{sec:unification}

According to the Unified model of Active Galactic Nuclei (AGN), the
central source has fundamentally same properties in the two types of
AGN \citep[e.g.,][]{Antonucci85,Antonucci93,UrryPadovani95}.  It is
only due to an obscuring medium, presumed to be a geometrically thick
torus, that the type 1 and the type 2 AGN differ.  In type 2 AGN, the
torus is assumed to be oriented edge-on to the observer to block the
view towards the Super Massive Black Hole (SMBH). In type 1 AGN, on
the other hand, the torus is oriented face-on, allowing a direct view
of the source. Observationally, there is a plenty of robust evidence
confirming this simple geometrical model in frequencies ranging from
the infrared (IR) to hard X-rays
\citep[][]{Antonucci85,Maiolino95,UrryPadovani95,Bassani99,Risaliti99,Lutz04,Heckman05,Buchanan06,Melendez08}.

The optical/UV luminosity of AGN is absorbed in the torus by the dust
grains and is re-radiated in the IR and NIR frequencies
\citep[e.g.,][]{Rees69,EdelsonMalkan86,Barvainis87,Pier92,LaorDraine93}.
This bright IR emission has been studied with interferometry in the
recent years \citep[e.g.,][]{Jaffe04,Packham05,Prieto05} allowing one
to estimate the torii physical sizes. These turned out to be rather
small, ranging from $\sim 0.03$ to a pc in NIR
\citep[e.g.,][]{KishimotoEtal11} frequencies to pc to tens of pc in
$12 \mu m$ \citep{TristramEtal09,Tristram11}.

Despite all the solid evidence for the existence of torii, no
convincing theoretical model has been ever produced to explain the
torus properties. The underlying theoretical difficulty is the
following. To harbor dust, gas needs to be not hotter than $\sim
2000$ K. The sound speed of such gas is only $\sim $ km s$^{-1}$,
whereas the vertical extent of the torus requires sound velocities of
at least a few hundred km s$^{-1}$. The torus could also be made of
cool clouds with large random velocities supporting the torus from
collapse. However, in order to provide a large enough absorbing column
depth, there should be $N_{\rm los}\sim $ several to ten clouds on a
typical line of sight. But this also means that these clouds collide
$\sim 2 \pi N_{\rm los}\gg 1 $ times per orbit
\citep{Krolik86,KB88}. How Mach number $\simgt 100$ collisions do not
destroy the gas clouds completely in a fraction of a period, given
their frequent collisions, appears to be beyond common sense
\citep[to solve this dilemma,][proposed the clouds to be very strongly
  magnetized]{Krolik86}.

Having said this, we note that a single universal model for AGN obscuration
may not even exist as the dominant absorption mechanism may vary from source
to source and even in time in the same source \citep{Risaliti02}. Examples of
processes likely to provide obscuration in AGN are: winds driven by a variety
of processes
\citep[e.g.,][]{Konigl94,Kartje99,Elvis00,Elvis04,Elitzur05,Proga03c,NC07},
warped accretion discs \citep{Nayakshin05}, clumpy medium produced by star
formation \citep{Wada02,SchartmannEtal10}, infra-red radiation supported torii
\citep{Krolik07}, temporary obscuration events produced by Broad Line Region
(BLR) clouds perhaps in a cometary shape \citep{MaiolinoEtal10}, and of course
obscuration by larger scale structures (the dust lanes, etc.) in the host
galaxy.

\subsection{Constraints on the smaller grains}\label{sec:small} 

We now consider constraints on the population of the smallest grains of a
single radius, $a$.

\subsubsection{Absorption by the grains in the optical}\label{sec:optical}

Absorption cross section, $\sigma_{\rm abs}$, for grains with size $a$ larger
than $0.1-1\mu$ \citep[depending on grain composition; see figures 2-4
  in][]{LaorDraine93} is approximately equal to their geometric area, $\pi
a^2$ in wavelengths from optical to soft X-rays ($h\nu \simlt 1 keV$). Within
our order of magnitude treatment, this provides a sufficiently accurate
prescription.

Let the number density of grains inside the absorber be $n_a$. The number of
grains on a line of sight is
\begin{equation}
N_a = n_a \pi a^2 R\;,
\label{nlos_a}
\end{equation}
where $R$ is the radial thickness of the torus, which we consider to
be of the
order of the distance to the SMBH. We note that $N_a \gg 1$ or else, due to
Poisson statistics, a sizable fraction of the lines of sight would contain no
dust grains at all. This would contradict the strong absorption of the broad
lines in the type 2 AGN \citep{Antonucci85}.

\subsubsection{Constraints from the X-ray frequencies}\label{sec:x}

Individual grains with size $a < a_0\simlt 1000\; \mu$m are optically
thin to X-rays with energy of a few keV (X-rays with this
characteristic energy typically determine the absorption column depth
in observations). Therefore, in contrast to the optical frequencies,
absorption of the primary AGN continuum in the X-rays measures the
total column depth of the grains on the line of sight. This column
depth can be related to the total mass of the small grains for a given
size of the torus.  If the volume of the torus is $V_t \approx 2\pi
R^3/3$, then the total mass of the grains in the torus is
\begin{equation}
M_a = {2\pi R^3 \over 3} n_a \rho_a {4\pi \over 3}a^3\;,
\label{Ma}
\end{equation}
where $\rho_a$ is the material density of a grain. The column depth of the
grains on a line of sight is
\begin{equation}
\Sigma_a = n_a R {4\pi \over 3}\rho_a a^3\;.
\label{sigma_a}
\end{equation}
The required surface density of metals in AGN absorbers can be deduced
from the ``hydrogen column'' $N_H$ reported in X-ray surveys
\citep[e.g.,][]{Sazonov04,SazonovEtal07,Guainazzi05}. These assume
that the metal-to-gas mass ratio is $\zeta_{\rm met} = 0.02$ as
appropriate to the gas of Solar composition. Our model then ought to
satisfy $\Sigma_a = \zeta_{\rm met} N_H m_H$, where $m_H$ is the mass
of hydrogen atom. This gives a constraint on the product
\begin{equation}
a N_a = {3 \zeta_{\rm met} N_H m_H \over 4 \rho_a} \approx 10^{-3} N_{23}\;,
\label{a_N_a}
\end{equation}
in cm, where we set $\rho_a = 2$ g cm$^{-3}$ and $N_H = N_{23} 10^{23}$
cm$^{-2}$. The total mass of the grains (equation \ref{Ma}), required to
fulfill the X-ray constraints, becomes
\begin{equation}
M_a = {2\pi \over 3}R^2 \zeta_{\rm met}N_H m_H = 300\msun\; R_{pc}^2 N_{23}
\label{Ma_x}
\end{equation}
For a given number of grains on the line of sight, $N_a$, the grain
size can be estimated:
\begin{equation}
a = 10 \mu m \; N_{23} N_a^{-1}\;.
\label{a_x}
\end{equation}
Note that if $N_a \gg 1$ then $a\simlt 1\mu$m for $N_{23}=1$.

\subsubsection{The role of gas}\label{sec:gas}

In contrast to debris disc systems where gas is presumed to be absent
\citep{Wyatt08}, the AGN torii cannot be totally devoid of gas because AGN is
fed by accretion of gas. We can estimate the minimum column density of gas in
the torus, $\Sigma_g = \rho R$, where $\rho$ is gas density in the torus via
the mass continuity for gas accreting through the torus:
\begin{equation}
\Sigma_g \sim {\dot M \over 2 \pi R v_{\rm ff}} = 5\times 10^{-4} \;\hbox{g
  cm}^{-2}\; \dot M_{-2}
R_{pc}^{-1/2} M_8^{-1/2}\;,
\label{gas}
\end{equation}
where $\dot M_{-2} = \dot M/(0.01 \msun/$yr) is the dimensionless gas
accretion rate onto the AGN, $R_{pc} = R/(1$pc), and $v_{\rm ff} =
(2G\mbh/R)^{1/2} = 2^{1/2} v_K$ is the free-fall velocity.

Aerodynamic drag force on the grains due to the presence of gas is
likely to be significant. For supersonic grain speeds, $v$, the drag
force is $F_d \approx (1/2) \pi a^2 \rho v^2$ \citep{WA61}. The
stopping time of the grain, $t_s = (4\pi/3)\rho_a a^3
v/F_d$. Comparing this to dynamical time, $t_{\rm dyn} = R/v_K$, we
have $t_s/t_{\rm dyn} = (8/3) a\rho_a /\Sigma_g$. Thus grains of
micron size will be stopped by the aerodynamic friction in a dynamical
time even at the minimum $\Sigma_g$ calculated above. Grains of $\sim$
cm sizes will be strongly affected by the gas after a sufficiently
long time.

This demonstrates that small grains that are crucial to the obscuration
schemes of AGN are ``frozen in'' with the gas. Therefore, the issue of the
vertical pressure support for the torus support must be addressed in this
model even if the origin of the small grains is a vertically extended
collisional cascade.

\subsubsection{Radiation pressure from the AGN}

Radiation pressure of the AGN can blow out the grains away if the bolometric
luminosity, $L_{\rm bol} = l_{\rm bol} \ledd$, is large enough. Here $L_{\rm
  Edd} = 4\pi G \mbh c/\kappa_{\rm e}$ is the Eddington luminosity of the black
hole, $\kappa_{\rm e}$ is the electron scattering opacity. In the case of an optically
thin torus, \cite{LaorDraine93} find that grains smaller than
\begin{equation}
a_{\rm b} \approx 6\times 10^3 \mu m\; l_{\rm bol}
\label{ablow}
\end{equation}
are blown out by the radiation field. Clearly, except for very dim
sources where $l_{\rm bol} < 10^{-4}$, $\mu$m sized grains are
expected to be driven out quickly if the torus is optically thin.

However, the absorber we are interested in is optically thick to most
of the AGN radiation (dominated by the UV bump). We expect the
radiation pressure incident on the AGN-faced side of the absorber to
be $\sim L_{\rm bol}/2c$, where $1/2$ comes from the fact that roughly
a half of the AGN radiation is intersected by the torus. This factor
should be reduced further if AGN radiation is beamed in the direction
perpendicular to the absorber's symmetry plane. The absorber's weight
is $G\mbh M_t/R^2$, where $M_t$ is the total torus mass, consisting of
the small grains mass, $M_a$, and the gas mass, $M_g$. Requiring that
the torus weight is greater than the radiation pressure on it, we find
that
\begin{equation}
M_t > 2\pi R^2 l_{\rm bol} \kappa_{\rm e}^{-1} \approx 10^3 \msun l_{-2} R_{pc}^2 \;,
\label{Ma_min}
\end{equation}
where $l_{-2} = l_{\rm bol}/0.01$. Since $M_a =
(8\pi/9)R^2 \rho_a a N_a$, we get a radius independent constraint on the
product $a N_a$:
\begin{equation}
a N_a > {9 M_a\over 4\kappa_{\rm e} \rho_a M_t} l_{\rm bol} \approx 3 l_{\rm
  bol} {M_a \over M_t} =  3 l_{\rm
  bol} f_d\;,
\label{an_min}
\end{equation}
where we introduced the dust mass fraction in the torus, $f_d \equiv M_a/(M_g
+ M_a)$ for brevity. If the torus is composed of gas with the usual
dust-to-gas abundance then $f_d = 0.01$. If the torus is dust free, like a
classical debris disc around a star, then $f_d = 1$.

Interestingly, equation \ref{a_N_a} now requires that the hydrogen column depth
of the absorber exceeds
\begin{equation}
N_{23} > 30 f_d \,l_{-2}\;.
\label{n23_min}
\end{equation}
This equation shows that, even for the normal dust-to-gas mass ratio
of $f_d=0.01$, {\em static} AGN torii would have to be Compton-thick
($N_{23} > 10$) in bright Eddington-limited quasars, for which
$l_{-2}\sim 100$. Static Compton-thin torii are not feasible for
quasars: Compton-thin torii must be in the state of an outflow driven
by the quasar's radiation pressure.

\subsubsection{Internal pressure support for grains}\label{sec:internal}

We shall now argue that radiation released in the torus internally by
accretion of gas through an accretion disc or by the stars may be sufficient
to inflate the torus vertically. This aspect of our model is therefore similar
to the model of \cite{Krolik07,SK08}, and is also related to larger scale disc
models of \cite{Thompson05}. The radiation pressure in the middle of the torus
can be estimated as
\begin{equation}
P_{\rm rad} \sim \tau_t {F_{\rm t} \over c}\;,
\label{prad}
\end{equation}
where $F_{\rm t}$ is the radiation flux emerging from the torus due to the
internal energy liberation, and $\tau_t = \kappa_a \Sigma_a\gg 1$ is the
optical depth of the torus. We assume that grains are sufficiently large to be
in the geometric optics absorption regime for the infrared radiation of the
torus as well, so that $\kappa_a = (\pi a^2)/(4\pi \rho_a a^3/3) = 3/(4\rho_a
a)$.  As the torus vertical scale height is about its radius, the pressure
balance in the vertical direction reads $P_{\rm rad} \simeq
(G\mbh/2R^2)\Sigma_t$. Thus, $F_{\rm t} \simeq (G\mbh c/2\kappa_aR^2)
f_d^{-1}$. Given that the absorber's surface area is $\sim 2 \pi R^2$, the
internal radiation flux can be converted into the luminosity of the torus
generated internally:
\begin{equation}
{L_{\rm t} \over \ledd} \simeq  {\kappa_{\rm e} \over 4 f_d \kappa_a}\approx
5 \times 10^{-6} \; f_d^{-1}\; a_{\mu m}\;,
\label{Lt}
\end{equation}
where $a_{\mu m} = a/(1 \mu$m). 

This estimate gives the internal luminosity of the torus needed to
provide the vertical pressure support. Is this a reasonable luminosity
to be expected from stars within the torus? To answer this, we assume
that the luminosity of the stellar population with mass $M_*$ is
dominated by young massive stars (anticipating that AGN activity and
star formation in the central parsec go hand in hand). Integrating
over the standard stellar mass function, we get $L_* \approx 0.002 \;
(4\pi G c M_{*}/\kappa_{\rm e})$, where $M_*$ is the total mass of the
young stars. The expression in the brackets is the Eddington
luminosity for mass $M_{*}$. If the mass function of stars is
top-heavy, as in the Galactic Centre star formation event inside the
central parsec \citep{NS05,PaumardEtal06,BartkoEtal10}, the luminosity
per unit mass can be higher, which we parametrize by introducing a
dimensionless factor $\epsilon_*$ which is greater than unity for a
top-heavy IMF.  Therefore, to satisfy equation \ref{Lt}, the total
mass of young stars inside the absorber must equal
\begin{equation}
M_* \approx 0.005 \mbh  \; {1 \over f_d \epsilon_*} \;  a_{\mu m}\;.
\label{mstar}
\end{equation}
Unless $f_d \epsilon_* \ll 1$, this number does not appear to be
excessively large: it is comparable with the gaseous mass of a
marginally unstable AGN accretion disc \citep{NC05} that is needed to
initiate star formation in an AGN disc. Finally, for reference we note
that for a stellar population composed of Sun-like stars only,
$\epsilon_* \approx 0.02$, whereas for an extremely top-heavy IMF,
$\epsilon_*$ can be as large as $\approx 500$.

\subsubsection{The inner edge of the torus}\label{sec:inner_edge}

Dust grains heated by the AGN to temperature greater than $\sim 1000$ K are
sublimated. The effective temperature of the radiation from an AGN with
luminosity $L_{\rm bol}$ a distance $R$ away is found from $T_{\rm eff}^4 \approx
L/(4\pi R^2 \sigma_B)$. This defines the inner edge to any dust-dominated
torus,
\begin{equation}
R_{pc} \approx T_3^{-2} l_{\rm bol}^{1/2} M_8^{1/2}\;,
\label{rpc}
\end{equation}
where $T_3 = T_{\rm eff}/10^3$ K is the grain sublimation temperature in units
of $10^3$ K.  We glossed over the fact that different species of dust and
different size dust can be sublimated at slightly different temperatures
\citep[e.g.,][]{LaorDraine93}.

\subsection{Summary of observational constraints}\label{sec:sum_const}

Large solids feeding the cascade can be said to be ``adiabatically''
collisionless, e.g., not likely to suffer a serious collision in a dynamical
time (eq. \ref{tcoll3}). This is an attractive feature of our model: large
asteroids can remain in a kinematically and geometrically thick distribution
for astrophysically interesting times, as required by AGN obscuration models
\citep{Antonucci85,Krolik86,KB88}. 

However, the small grains are strongly influenced by even modest
amounts of gas present in the torus (\S \ref{sec:gas}). Further, if
small grains form an optically thick system they should also be
collisional if moving at large random velocities \citep[the arguments
  of][apply in this case to the individual dust grains]{Krolik86}. The
upshot of this is that the small grains should settle dynamically to a
symmetry plane (which may be in general warped). Therefore, a
radiation pressure support internal to the torus is still needed to
keep the torus geometrically thick.

On the other hand, the observed obscuration in the optical and X-ray
frequencies inside the AGN absorbers gave us a minimum column depth in the
grains that the screen of the small grains needs to possess in order to
account for the absorption. This translates into the total mass of the grains
if the size of the torus is known (see equation \ref{Ma_x}). The size of the
torus is a fraction of a parsec based on both interferometry observations and
the dust sublimation constrains (equation \ref{rpc}).

We then assumed that the radiation pressure support comes from the same stars
that formed the dust, and estimated the total mass in the stars in equation
\ref{mstar}. We can now close the logical loop by checking if these stars
could actually provide the solids needed to fuel the cascade in the first
place.

Take the ratio of the torus mass in the grains, obtained by using X-ray
constraints (equation \ref{Ma_x}), to that of the stars needed for the
pressure support:
\begin{equation}
{M_a \over M_*} = 5 \times 10^{-4} {R_{pc}^2 N_{23} \over M_8 \, a_{\mu
    m}}\; {\epsilon_* f_d}\;.
\label{Ma_Ms}
\end{equation}
Now, we can eliminate $R_{pc}$ from this equation by assuming that the torus
is larger than the dust sublimation radius given by equation
\ref{rpc}. Finally, in order for the smallest grains to withstand the
radiation pressure from the AGN, we required a minimum column depth $N_{23}$
in equation \ref{n23_min}. We now use that in equation \ref{Ma_Ms} and find
that the ratio of the mass in the small grains to that of the stars should
satisfy
\begin{equation}
{M_a \over M_*} > 1.5 \times 10^{-6} \frac{\epsilon_* l_{-2}^2 f_{-1}^2}{a_{\mu
    m} T_3^4}\;,
\label{Ma_Ms2}
\end{equation}
where $f_{-1} = f_d/0.1$.  This is a significant constraint on our model. It
delineates the regimes where the fragmentation cascades may or may not be
important for the observed AGN obscuration.

The realistically available budget of solids has an obvious upper limit -- the
total mass of metals with respect to that of H and He, i.e., $M_a/M_* < M_Z/M_*
\sim 0.01$. This is of course wildly optimistic. To do better than that,
consider observations of planetary systems and their debris discs. In the
Solar System, most of the solid mass, outside of the Sun, is in the solid
cores of the giant planets. That amounts to $\sim 50-60 \mearth$, which gives
a fraction $M_Z/\msun\approx 1.5 \times 10^{-4}$ of the total system's mass.
Proto-planetary discs, before turning into debris discs, are found to contain
up to a few to ten times more mass in the dust than this \citep[e.g., see
  Fig. 3 in][]{Wyatt08}. Therefore, even being optimistic, it is hard to
expect that solids in asteroids, terrestrial like planets and solid cores of
giant planets account for more than $\sim 10^{-3}$ of the stellar
mass. Therefore, in parameter space where equation \ref{Ma_Ms2} requires more
small grains than $\sim 10^{-3}$ of the stellar mass the model does not appear
realistic.

\section{Discussion}\label{sec:discussion}

In this paper we argued that it is quite likely that planets may form around
SMBH in self-gravitating $\simlt$ parsec-scale discs as a compliment to the
star formation process that is known to occur there. We also argued that the
geometrical arrangement of different rings of stars and planets may be very
mis-aligned and should thus lead to very energetic collisions between solids
from different stellar discs. This is certain to fuel fragmentation cascades
similar to those occurring around stars \citep{Wyatt08} except for the much
more extreme collision velocities. We found (\S \ref{sec:fragmentation}) that
smaller solid bodies, e.g., smaller than $\sim$ hundreds of km in diameter are
rapidly destroyed in catastrophic collisions. Solid objects larger than $\sim
1000$ km in diameter should feed fragmentation cascades for astrophysically
interesting time scales, e.g., millions of years (equation \ref{tcoll3}). We
then proceeded to consider whether small grains, presumably the end product of
the fragmentation cascade, can explain the observed AGN obscuration (\S
\ref{sec:super}). We shall now collect the physical constraints we obtained
there in order to make observational predictions and thus test the model
further.

\subsection{No obscuring torus near Sgr A*}\label{sec:sgra}

The best observed SMBH is \sgra\ in the centre of our Galaxy, as
discussed in the Introduction. The column density of absorbing
material on the line of sight to \sgra\ is $N_H\simlt 10^{23}$
\citep{Baganoff03}, and it appears that most of that is outside the
central parsec of the Milky Way (e.g., \citealt{Muno04}).
Far infrared observations show that the
absorbing column density of the central $\sim 2$ pc region near
\sgra\ is $N_H \simlt 10^{22}$ \citep{Zylka95,EtxaluzeEtal11}.

At the same time we know that there was a star formation event $\sim
6$ Myr ago, with the total stellar mass less than $M_t \sim 10^4\msun$
\citep{PaumardEtal06}, and that the geometrical arrangement of the
stars is far from flat \citep{BartkoEtal09}. If these massive
\citep{BartkoEtal10} stars also came with asteroids and comets then a
fragmentation cascade is expected.

However we argue that \sgra\ star formation event simply did not make
enough stars to create a significant fragmentation cascade. First,
assuming generously that the total mass of ``new'' solid material is
$M_Z = 10^{-4} M_t \simlt 1 \msun$, we estimate (cf. \S \ref{sec:x})
the maximum absorbing column density of these solids if spread around
as fine grains: $N_H \sim M_z/(2\pi R^2 m_H \zeta_{\rm met}) \approx
10^{22}$~cm$^{-2}$ for $R=0.3$ pc. This is comparable to the deduced
$N_H$ in the central parsec \citep{EtxaluzeEtal11}. Furthermore, the
collisional time scale for the cascade to develop for \sgra\ stars is
$\sim 2\times 10^9$ yr (cf. equation \ref{tcoll3} for the parameters
used above and $D_8=1$). Physically, such a tenuous cascade is unable
to engage the largest bodies in which most of the solid mass is stored
(at least in the Solar System). Furthermore, the mass spectrum of the
young ``disc'' of stars in the central pc of the Milky Way is
unusually top heavy, with a at least an order of magnitude deficit of
Solar type stars \citep{NS05,BartkoEtal10}, and therefore it is not
clear whether asteroids and comets managed to form in the usual
numbers around these massive stars.

We thus conclude that it is very unlikely that a rather limited star
formation event that occur ed in the central parsec of the Milky Way,
$M_t \simlt 10^4 \msun$, could have produced enough solids to make a
detectable obscuring torus.

\subsection{No obscuring cascades at low accretion rate AGN}

The example of \sgra\ considered above demonstrates that isolated and
limited star formation episodes inside the central parsecs of AGN may
be insufficient (even under most generous assumptions) to provide a
significant enough amount of solids to initiate significant
fragmentation cascades.  We shall estimate the minimum total mass in
the circum-nuclear AGN star formation event(s) that could provide
enough solids for a given obscuring column density $N_H$.

The torii for low luminosity objects are expected to be physically
small, e.g., $R \simlt 0.1$ pc \citep[the torus size seems to be
  consistent with the location where grains would sublimate, e.g., see
  Fig 4 in][]{KishimotoEtal11}. The total mass of small grains needed
to account for a given $N_H$ is given by equation \ref{Ma_x} as $M_a
\sim 3\msun (R/0.1\mbox{ pc})^2 N_{23}$. The minimum mass in comets
and asteroids, $M_Z$, should be at least this large, obviously, so
$M_Z \gg M_a$. If $M_Z = \delta_Z M_t = 10^{-4} \delta_{-4} M_t$, then
\begin{equation}
M_t \gg 3 \times
10^{4}\msun (R/0.1\mbox{ pc})^2 N_{23}/\delta_{-4}\;.
\label{mt_low}
\end{equation}
We conclude that, realistically, a star formation event with at least
$10^5-10^6\msun$ of stars is required to make an obscuring torus. It
is theoretically uncertain how much gas is accreted onto the SMBH
during such star formation events in the inner parsec
\citep[see][]{NayakshinEtal07,ZubovasEtal11}. However, if we assume
that the AGN activity phase lasted for $\sim 10^7$ yrs, and that only
a small fraction of the gas was accreted onto the SMBH, then the
average accretion rate is $\dot M\ll 0.01 -0.1 \msun/$yr.

Further, at a low enough accretion rate in the AGN disc the star
formation turns off completely because the disc is no longer
self-gravitating \citep[e.g.,][]{Paczynski78,Collin99}. Consider the
case $M_8=1$ as an example. Figure 1 in \cite{Goodman03} shows that
independently of the viscosity prescription in the disc, the innermost
$0.01 -0.1$ pc are gravitationally stable for accretion rates lower
than about $10^{-4} \msun$ yr$^{-1}$. Therefore, no star or
comet/asteroid formation is expected at such low accretion rates.

If the scaling between an average accretion rate in the disc on sub-pc
scales and the AGN accretion rate were clear, this could be mapped
into a luminosity-dependent prediction for AGN obscuration.  However,
one should of course keep in mind the caveat that AGN luminosity could
vary significantly in time on time scales shorter than those required
for a significant evolution of the collisional cascades. Thus, there
could be sources that are presently in the LLAGN state but were much
more active in a recent past, and still have enough solid debris
needed for the AGN obscuration cascade.

\subsection{No obscuring cascades in quasars}\label{sec:quasars}

Similarly, we make a broad brush definition of a quasar as an AGN accreting
gas at nearly the Eddington accretion rate, setting $l_{\rm bol}\sim 1$. In
this case, $l_{-2} = 100 l_{\rm bol} \sim 100$, and equation \ref{Ma_Ms2}
demands an unreasonably high fractional mass in the small grains:
\begin{equation}
{M_a \over M_*} > 1.5 \times 10^{-2} \frac{\epsilon_* l_{\rm bol}^2 f_{-1}^2}{a_{\mu
    m} T_3^4}\;.
\label{Ma_Msq}
\end{equation}
Even for the Solar abundance gas, for which $f_{-1} = f_d/0.1 =
0.1$, the requirements are still not very comfortable. Further, $M_a$ is the
total mass of the small grains. Since these are fed by the fragmenting cascade
of the larger bodies, we would expect the mass in the solid bodies to be much
higher than that in the small dust. This would correspondingly increase the
required mass of solids in equation \ref{Ma_Msq}.

Physically, the inability of our model to provide obscuration for
Eddington-limited sources is due to the large radiation pressure of
the quasar in such sources. This requires the torus to be very massive
(equation \ref{Ma_min}), which is unreasonably high if the torus is
made by the fragmentation cascade.

\subsection{Torii in intermediate luminosity AGN:
  physically small}

We find that the intermediate luminosity AGN, which we operationally
define as $l_{\rm bol} \sim 0.01$, appear to be able to both host a
starburst to make the solids for the cascade, and also make enough
stars to provide the internal pressure support. The fraction of solids
that should be put into asteroids does not appear excessive in this
case either (equation \ref{Ma_Ms2}).

One interesting aspect of the model, however, is that the resulting obscuring
torii must be relatively small. Rescaling equation \ref{rpc}, we have
\begin{equation}
R_{pc} \approx 0.1 T_3^{-2} l_{-2}^{1/2} M_8^{1/2}\;.
\label{rpcq}
\end{equation}
Such small torii sizes are reasonably consistent with observations. It is
interesting to note that if we were to appeal to a torus much larger than
this, say by an order of magnitude, then we would again run into the mass
budget problem for solids (cf. equation \ref{Ma_Ms}).

Note that optically thin ``torii'' of larger physical extent could
however be made by some fraction of the grains lost to an outflow from
the AGN.

\subsection{Large dust in AGN torii}

The silicate spectral feature near 9.7 $\mu m$ is surprisingly weak in
typical AGN spectra (e.g., \citealt{Hao07,Shi06}). In
addition, there is also a discrepancy in the 
inferred column density of neutral hydrogen via observations in the
optical and via X-rays. A number of authors argued that these
peculiarities of AGN absorbers point to ``anomalous'' properties of
the AGN dust with regard to Galactic dust, and suggested that the AGN
dust particles may be larger, e.g., $a \simgt 1-10 \mu m$
\citep[e.g.,][]{LaorDraine93,BrandtEtal96,MaiolinoEtal01a,MaiolinoEtal01b}.
Our model is not detailed enough for us to be able to predict the size
distribution of smaller grains and to compare directly with the
``peculiar'' AGN dust. However, simply the fact that the dust in our
model results from the fragmentation cascade of larger bodies suggests
that such dust should be physically larger than its Galactic
counterpart. Therefore, large dust in AGN torii, if confirmed with
future observations, could originate in the collisional cascade
discussed here. Small dust may be also present, if brought in from
outside by gas inflow, as argued in \S \ref{sec:gas}, or if
fragmentation cascade continues to very small, $a\ll \mu$m, scales.

\section{Conclusions}

We suggested that solid bodies, from asteroids and comets to large
planetary cores, may form in AGN accretion discs due to gravitational
instability of the latter. These star and planet formation episodes
are likely to occur in randomly oriented planes and thus result in a
quasi-spherical distribution of stars and solids. The solids from
different discs must then collide at velocities of order $1000$
km~s$^{-1}$, leading to fragmentation cascades that can split even the
largest bodies.

We noted that such cascades are likely to grind the solids all the way
into the microscopic dust. The small dust particles must absorb and
re-radiate the AGN luminosity efficiently. This may be relevant to the
well known but still poorly understood AGN obscuration. We attempted
to put various physical constraints on this picture. We found that
such asteroid-fed torii may only work for relatively mildly bright
AGN, whereas in AGN approaching their Eddington limits the dust would
be driven away by the radiation pressure. Also, the least luminous
AGN, such as \sgra, are unlikely to host strong enough
starbursts. Solid bodies in this case are not numerous enough to fuel
strong fragmentation cascades that could produce an observationally
significant amount of microscopic dust.

We should also point out that the fragmentation cascades discussed
here may co-exist with ``conventional'' gas-rich AGN torii
\citep{KL89} made by other means, such as outflows or inflows of gas
or supernova explosions. If these are not co-spatial with the cloud of
solid bodies considered here, then they are independent of each
other. If the conventional torus is co-spatial with the fragmentation
cascade then the cascade may be an additional source of dust
particles.


\section{Acknowledgments}

The authors happy to acknowledge helpful comments on the manuscript by Mark
Wyatt. Theoretical astrophysics research in Leicester is supported by an STFC
Rolling Grant. SN and SS thank Max Planck Institute where this research was
completed for hospitality. The research made use of grants RFBR 09-02-00867a
and NSh-5069.2010.2 and programs P-19 and OFN-16 of the Russian Academy of
Sciences. SS acknowledges the support of the Dynasty Foundation.


\end{document}